\documentclass{article}
\usepackage{amssymb}
\usepackage{makeidx}
\usepackage{amsmath}

\input{tcilatex}

\begin{document}

\title{Prisoners' Dilemma in Presence of Collective Dephasing}
\author{Ahmad Nawaz\thanks{%
email: ahmad@ele.qau.edu.pk} \\
National Centre for Physics, Quaid-i-Azam University Campus, \\
Islamabad, Pakistan.}
\maketitle

\begin{abstract}
We quantize prisoner dilemma in presence of collective dephasing with
dephasing rate $\gamma $. It is shown that for two parameters set of
strategies $Q\otimes Q$ is Nash equilibrium below a cut-off value of time.
Beyond this cut-off it bifurcates into two new Nash equilibria $Q\otimes D$
and $D\otimes Q$. Furthermore for maximum value of decoherence \ $C\otimes D$
and $D\otimes C$ also become Nash equilibria. At this stage the game has
four Nash equilibria. On the other hand for three parameters set of
strategies there is no pure strategy Nash equilibrium however there is mixed
strategy (non unique) Nash equilibrium that is not affected by collective
dephasing.
\end{abstract}

\section{Introduction}

In game theory Nash equilibrium (NE) is central solution concept. It is a
set of strategies from which unilateral deviation of any player reduces
his/her payoff. However, some shortcomings are also associated with this
solution concept. First, it is not necessarily true for each game to have
unique NE.\ Battle of Sexes (BoS) and Chicken game (CG) are well known
examples in this regard. Second, in some cases NE could result in an outcome
that is far from the benefit of the players. Prisoners' Dilemma (PD) is an
example where the rational reasoning forces the players to\ fall in a
dilemma with worst outcomes. Quantum game theory helps resolve such dilemmas 
\cite{eisert,marinatto} and shows that quantum strategies can be
advantageous\ over classical strategies \cite{eisert,meyer,flitney}. One of
the foremost and elegant step in this direction is by Eisert \textit{et al} 
\cite{eisert} to remove the dilemma in PD. In this quantization scheme the
strategy space of the players is a two parameters set of $2\times 2$ unitary
matrices. Starting with maximally entangled initial quantum state the
authors showed that for a suitable quantum strategy the dilemma disappears
from the game. They also pointed out a quantum strategy which always wins
over all the classical strategies. Later on, Marinatto and Weber \cite%
{marinatto} introduced another interesting and simple scheme for the
analysis of non-zero sum games in quantum domain. They gave Hilbert
structure to the strategic spaces of the players. They also used maximally
entangled initial state and allowed the players to play their tactics by
applying probabilistic choice of unitary operators. Applying their scheme to
an interesting game of BoS they found the strategy for which both the
players can achieve equal payoffs.\ For both these schemes role of initial
quantum state remained important \cite%
{flitney,azhar,azhar1,azhar2,jiang,rosero}. In our earlier work on the
subject, we introduced a generalized quantization scheme for two person
non-zero sum games that gives a relationship between these two apparently
different quantization schemes \cite{nawaz}. Separate set of parameters were
identified for which this scheme reduces to that of Marinatto and Weber \cite%
{marinatto} and Eisert \textit{et al} \cite{eisert} quantization schemes.
Furthermore some other interesting situations were identified which were not
apparent within the existing two quantizations schemes.

Players have to share qubits to play quantum games and the qubits are prone
to decoherence. Many authors obtained interesting results by quantizing
games in presence of noise \cite%
{johnson,chen,flitney,flitney-3,nawaz-correlated,ramzan}. Chen \textit{et al 
}\cite{chen} analyzed PD in the presence of three prototype quantum channels
and showed that the payoffs gradually decrease with increasing noise without
affecting the NE of the game. Later on Flitney and Abbott \cite%
{flitney,flitney-3} studied quantum games in presence of decoherence to find
the advantage that a quantum player could have over a classical one. In
their scheme this advantage of quantum player is termed as the measure of
`quantum-ness' of a quantum game subjected to decoherence. They showed that
the advantage of quantum player reduces with increasing decoherence and at
maximum value of decoherence it disappears completely. In a $2\times 2$
symmetric game for the maximum value of decoherence the payoffs of both
players become same and at this stage the classical game cannot be
reproduced. Then, Nawaz and Toor \cite{nawaz-correlated} analyzed quantum
games in presence of correlated noise. They also restricted one of the
players to play classical strategies. They showed that the effects of memory
and decoherence become effective only when game starts from a maximally
entangled state and the measurement is performed in entangled basis. In this
case the quantum player outperforms the classical one. They also highlighted
the fact that memory controls payoffs reduction due to decoherence and for
maximum value of memory decoherence becomes ineffective. Following the same
lines Ramzan \textit{et al }\cite{ramzan} quantized PD, BoS and CG in
presence of three prototype quantum correlated channels using generalized
quantization scheme. They also observed that the effects of the memory and
decoherence become effective for maximally entangled initial state and
entangled measurement for which the quantum player outperforms the classical
player. They also noticed that memory has no effect on NE. An important type
of noise\textrm{\ that has been studied extensively in quantum information
theory is collective dephasing \cite{li,keil,dong,lidar,Khodjasteh,han}}.%
\textrm{\ It plays crucial role in physical systems like trapped ions,
quantum dots and atoms inside a cavity.} It also allows the existence of the
decoherence free subspace \cite{zanardi}. In this paper we quantize PD in
presence of collective dephasing using our generalized quantization scheme
for games which allows us to perform measurements in entangled as well as in
product basis \cite{nawaz}. The game starts with a maximally entangled state
that has decohered by collective dephasing channel of dephasing rate $\gamma
.$ We show that for measurement in entangled basis when the players are
allowed to play the two parameters set of strategies then $Q\otimes Q$%
\textrm{\ which is the NE at }$t=0$ (no noise) \cite{eisert}\textrm{\
remains NE for }$e^{-2\gamma t}>\frac{1}{7}$. With increasing time $t$ when $%
e^{-2\gamma t}\leq \frac{1}{7}$ then $Q\otimes Q$ does not remain a NE but
there appear two new NE $Q\otimes D$\ and $D\otimes Q$ simultaneously.
Furthermore when $e^{-2\gamma t}\longrightarrow 0$ then besides $Q\otimes D$%
\ and $D\otimes Q$ the strategy pairs $C\otimes D$\ and $D\otimes C$ also
become NE. At this stage there are four NE in the game.\emph{\ }For the
measurement in product basis $C\otimes D$\ and $D\otimes C$ are decohence
free NE. For three parameters set of strategies, there exists no pure
strategy NE because for every strategy of one player, the other also has a
counter strategy. However there can be a mixed strategy (non unique) NE \cite%
{eisert1,benjamin} which also remains unaffected by collective dephasing
both for entangled and product basis measurements.

The paper is organized as follows: In section (\ref{prisoner}), after a
brief introduction to PD we quantize it in presence of collective dephasing
and section (\ref{con}) concludes the results.

\section{\label{prisoner}Quantization of PD with Collective Dephasing\textbf{%
\ }}

Prisoner dilemma is based on a story of two suspects, say them Alice and
Bob, who have allegedly committed a crime together. They have been arrested
and being interrogated in two separate cells. Each of suspects will have to
decide whether to confess the crime or to deny the crime without any
communication between them. In game theory the players decisions to confess
the crime is termed as to defect (strategy $D$) and to deny the crime is
called to cooperate (strategy $C$). This situation can be depicted in form
of a bimatrix shown below

\begin{equation}
\text{{\large Alice }}%
\begin{array}{c}
C \\ 
D%
\end{array}%
\overset{}{\overset{%
\begin{array}{c}
\text{{\large Bob}} \\ 
\begin{array}{cc}
C\text{ \ \ \ \ } & D%
\end{array}%
\end{array}%
}{\left[ 
\begin{array}{cc}
\left( 3,3\right) & \left( 0,5\right) \\ 
\left( 5,0\right) & \left( 1,1\right)%
\end{array}%
\right] }}.  \label{matrix-prisoner}
\end{equation}%
Depending upon their decisions the players get payoffs according to the
above payoff matrix. It is clear from the this payoff matrix that $D$\ is
the dominant strategy for both players. Therefore, rational reasoning forces
each of them to play $D$ resulting ($D,D$) as the NE with payoffs $\left(
1,1\right) ,$ which is not Pareto Optimal. However, it was possible for the
players to get better payoffs $\left( 3,3\right) $ if they would have played 
$C$\ instead of $D$. This is generally known as dilemma of this game.

For the quantum version of PD the classical strategies $C$\ (Cooperate) and $%
D$\ (Defect) are assigned two basis vectors $\left\vert C\right\rangle $\
and $\left\vert D\right\rangle $\ respectively, in a Hilbert space of a
two-level system. The state of the game at any instant is a vector in four
dimensional Hilbert space spanned by the basis vectors $\left\vert
CC\right\rangle ,$ $\left\vert CD\right\rangle ,$ $\left\vert
DC\right\rangle ,$ $\left\vert DD\right\rangle .$\ Here the entries in ket
refer to the qubits possessed by Alice and Bob respectively. But the qubits
are prone to decoherence due to their interaction with environment. For the
environment of the form of collective dephasing the state $\hat{\rho}$ of
system is transformed by the following master equation \cite{lidar}%
\begin{equation}
\frac{\partial \hat{\rho}}{\partial t}=\frac{\gamma }{2}\left( 2\hat{J}_{z}%
\hat{\rho}\hat{J}_{z}-\hat{J}_{z}^{2}\hat{\rho}-\hat{\rho}\hat{J}%
_{z}^{2}\right)  \label{collective dephasing}
\end{equation}%
where $\gamma $ is dephasing rate and $\hat{J}_{z}$ are the collective spin
operators defined as 
\begin{equation}
\hat{J}_{z}=\sum\limits_{i=1}^{2}\frac{\hat{\sigma}_{z}^{\left( i\right) }}{2%
}
\end{equation}%
with $\hat{\sigma}_{z}$ as the Pauli matrices. Under the action of
collective dephasing (\ref{collective dephasing}) the maximally entangled
state 
\begin{equation}
\left\vert \phi ^{+}\right\rangle =\frac{\left\vert 00\right\rangle
+i\left\vert 11\right\rangle }{\sqrt{2}}
\end{equation}%
shared by players becomes 
\begin{equation}
\rho \left( t\right) =\frac{1}{2}\left( 1-e^{-2\gamma t}\right) \left(
\left\vert 00\right\rangle \left\langle 00\right\vert +\left\vert
11\right\rangle \left\langle 11\right\vert \right) +\left\vert \phi
^{+}\right\rangle \left\langle \phi ^{+}\right\vert e^{-2\gamma t}.
\label{after dephasing}
\end{equation}%
In this case the game starts with the state (\ref{after dephasing}) that has
decohered by a collective dephasing channel of dephasing rate $\gamma $ for
time $t$. The players apply their strategies on this decohered state. The
strategy of the\ players is represented by the unitary operator $U_{i}$\
given as \cite{nawaz} 
\begin{equation}
U_{i}=\cos \frac{\theta _{i}}{2}R_{i}+\sin \frac{\theta _{i}}{2}C_{i},
\label{strategy}
\end{equation}%
where $i=A$\ or $B$\ and $R_{i}$, $C_{i}$\emph{\ }are the unitary operators
defined as%
\begin{align}
R_{i}\left\vert 0\right\rangle & =e^{i\phi _{i}}\left\vert 0\right\rangle ,%
\text{ \ \ }R_{i}\left\vert 1\right\rangle =e^{-i\phi _{i}}\left\vert
1\right\rangle ,  \notag \\
C_{i}\left\vert 0\right\rangle & =-\left\vert 1\right\rangle ,\text{ \ \ \ \
\ }C_{i}\left\vert 1\right\rangle =\left\vert 0\right\rangle .
\label{operators}
\end{align}%
\emph{\ }After the application of the strategies, the initial state given by
Eq. (\ref{after dephasing}) transforms into 
\begin{equation}
\rho _{f}=(U_{A}\otimes U_{B})\rho \left( t\right) (U_{A}\otimes
U_{B})^{\dagger }.  \label{final}
\end{equation}%
The payoff operators for Alice and Bob are

\begin{align}
P^{A}& =3P_{00}+P_{11}+5P_{10},  \notag \\
P^{B}& =3P_{00}+P_{11}+5P_{01},  \label{pay-operator}
\end{align}%
where 
\begin{subequations}
\label{oper a}
\begin{align}
P_{00}& =\left\vert \psi _{00}\right\rangle \left\langle \psi
_{00}\right\vert \text{, \ }\left\vert \psi _{00}\right\rangle =\cos \frac{%
\delta }{2}\left\vert 00\right\rangle +i\sin \frac{\delta }{2}\left\vert
11\right\rangle ,  \label{oper 1} \\
P_{11}& =\left\vert \psi _{11}\right\rangle \left\langle \psi
_{11}\right\vert ,\text{ \ }\left\vert \psi _{11}\right\rangle =\cos \frac{%
\delta }{2}\left\vert 11\right\rangle +i\sin \frac{\delta }{2}\left\vert
00\right\rangle ,  \label{oper 2} \\
P_{10}& =\left\vert \psi _{10}\right\rangle \left\langle \psi
_{10}\right\vert \text{, \ }\left\vert \psi _{10}\right\rangle =\cos \frac{%
\delta }{2}\left\vert 10\right\rangle -i\sin \frac{\delta }{2}\left\vert
01\right\rangle ,  \label{oper 3} \\
P_{01}& =\left\vert \psi _{01}\right\rangle \left\langle \psi
_{01}\right\vert \text{, \ }\left\vert \psi _{01}\right\rangle =\cos \frac{%
\delta }{2}\left\vert 01\right\rangle -i\sin \frac{\delta }{2}\left\vert
10\right\rangle ,  \label{oper 4}
\end{align}%
and\emph{\ }$\delta \in \left[ 0,\frac{\pi }{2}\right] $ is the entanglement
of measurement basis. These payoff operators reduce to that of Eisert's
scheme for $\delta =\frac{\pi }{2}$ \cite{eisert} and for $\delta =0$ these
transform into that of Marinatto and Weber's scheme \cite{marinatto}. The
payoffs for the players are found as 
\end{subequations}
\begin{eqnarray}
\$_{A}(\theta _{A},\phi _{A},\theta _{B},\phi _{B}) &=&\text{Tr}(P^{A}\rho
_{f})\text{,}  \notag \\
\$_{B}(\theta _{A},\phi _{A},\theta _{B},\phi _{B}) &=&\text{Tr}(P^{B}\rho
_{f})\text{,}  \label{payoff-generalized}
\end{eqnarray}%
where Tr represents the trace of a\emph{\ }matrix. Using Eqs. (\ref%
{matrix-prisoner}, \ref{after dephasing}, \ref{final}, \ref{pay-operator}, %
\ref{payoff-generalized}) the payoffs are obtained as 
\begin{eqnarray}
\$_{A}(\theta _{A},\phi _{A},\theta _{B},\phi _{B}) &=&\left[ 2+e^{-2\gamma
t}\cos 2\left( \phi _{A}+\phi _{B}\right) \sin \delta \right] \cos ^{2}\frac{%
\theta _{A}}{2}\cos ^{2}\frac{\theta _{B}}{2}  \notag \\
&&+\left[ 2-e^{-2\gamma t}\sin \delta \right] \sin ^{2}\frac{\theta _{A}}{2}%
\sin ^{2}\frac{\theta _{B}}{2}  \notag \\
&&+\frac{5}{2}\left[ 1-e^{-2\gamma t}\cos 2\phi _{A}\sin \delta \right] \cos
^{2}\frac{\theta _{A}}{2}\sin ^{2}\frac{\theta _{B}}{2}  \notag \\
&&+\frac{5}{2}\left[ 1+e^{-2\gamma t}\cos 2\phi _{B}\sin \delta \right] \sin
^{2}\frac{\theta _{A}}{2}\cos ^{2}\frac{\theta _{B}}{2}  \notag \\
&&-\frac{1}{4}\left[ e^{-2\gamma t}+2\sin \delta \right] \sin \theta
_{A}\sin \theta _{B}\sin \left( \phi _{A}+\phi _{B}\right)  \notag \\
&&-\frac{5}{4}\sin \theta _{A}\sin \theta _{B}\sin \left( \phi _{A}-\phi
_{B}\right) \sin \delta  \label{payoff-pd-a}
\end{eqnarray}%
\begin{eqnarray}
\$_{B}(\theta _{A},\phi _{A},\theta _{B},\phi _{B}) &=&\left[ 2+e^{-2\gamma
t}\cos 2\left( \phi _{A}+\phi _{B}\right) \sin \delta \right] \cos ^{2}\frac{%
\theta _{A}}{2}\cos ^{2}\frac{\theta _{B}}{2}  \notag \\
&&+\left[ 2-e^{-2\gamma t}\sin \delta \right] \sin ^{2}\frac{\theta _{A}}{2}%
\sin ^{2}\frac{\theta _{B}}{2}  \notag \\
&&+\frac{5}{2}\left[ 1+e^{-2\gamma t}\cos 2\phi _{A}\sin \delta \right] \cos
^{2}\frac{\theta _{A}}{2}\sin ^{2}\frac{\theta _{B}}{2}  \notag \\
&&+\frac{5}{2}\left[ 1-e^{-2\gamma t}\cos 2\phi _{B}\sin \delta \right] \sin
^{2}\frac{\theta _{A}}{2}\cos ^{2}\frac{\theta _{B}}{2}  \notag \\
&&-\frac{1}{4}\left[ e^{-2\gamma t}+2\sin \delta \right] \sin \theta
_{A}\sin \theta _{B}\sin \left( \phi _{A}+\phi _{B}\right)  \notag \\
&&-\frac{5}{4}\sin \theta _{A}\sin \theta _{B}\sin \left( \phi _{A}-\phi
_{B}\right) \sin \delta  \label{payoff-pd-b}
\end{eqnarray}%
These payoffs transform to that of Eisert \textit{et al} \cite{eisert} for $%
t=0$ and $\delta =\frac{\pi }{2}.$ As in our generalized quantization scheme
measurements can be performed in entangled ($\delta =\frac{\pi }{2}$) as
well as in product basis ($\delta =0)$. Therefore we take both the cases one
by one.

\subsection{Entangled Measurement}

For entangled measurement i.e. $\delta =\frac{\pi }{2}$ the payoffs given in
Eqs. (\ref{payoff-pd-a}) and (\ref{payoff-pd-b}) become%
\begin{eqnarray}
\$_{A}\left( \theta _{A},\phi _{A},\theta _{B},\phi _{B}\right) &=&\left[
2+e^{-2\gamma t}\cos 2\left( \phi _{A}+\phi _{B}\right) \right] \cos ^{2}%
\frac{\theta _{A}}{2}\cos ^{2}\frac{\theta _{B}}{2}  \notag \\
&&+\left[ 2-e^{-2\gamma t}\right] \sin ^{2}\frac{\theta _{A}}{2}\sin ^{2}%
\frac{\theta _{B}}{2}  \notag \\
&&+\frac{5}{2}\left[ 1-e^{-2\gamma t}\cos 2\phi _{A}\right] \cos ^{2}\frac{%
\theta _{A}}{2}\sin ^{2}\frac{\theta _{B}}{2}  \notag \\
&&+\frac{5}{2}\left[ 1+e^{-2\gamma t}\cos 2\phi _{B}\right] \sin ^{2}\frac{%
\theta _{A}}{2}\cos ^{2}\frac{\theta _{B}}{2}  \notag \\
&&-\frac{1}{4}\left[ 2+e^{-2\gamma t}\right] \sin \theta _{A}\sin \theta
_{B}\sin \left( \phi _{A}+\phi _{B}\right)  \notag \\
&&-\frac{5}{4}\sin \theta _{A}\sin \theta _{B}\sin \left( \phi _{A}-\phi
_{B}\right)  \label{payoff a}
\end{eqnarray}%
\begin{eqnarray}
\$_{B}\left( \theta _{A},\phi _{A},\theta _{B},\phi _{B}\right) &=&\left[
2+e^{-2\gamma t}\cos 2\left( \phi _{A}+\phi _{B}\right) \right] \cos ^{2}%
\frac{\theta _{A}}{2}\cos ^{2}\frac{\theta _{B}}{2}  \notag \\
&&+\left[ 2-e^{-2\gamma t}\right] \sin ^{2}\frac{\theta _{A}}{2}\sin ^{2}%
\frac{\theta _{B}}{2}  \notag \\
&&+\frac{5}{2}\left[ 1+e^{-2\gamma t}\cos 2\phi _{A}\right] \cos ^{2}\frac{%
\theta _{A}}{2}\sin ^{2}\frac{\theta _{B}}{2}  \notag \\
&&+\frac{5}{2}\left[ 1-e^{-2\gamma t}\cos 2\phi _{B}\right] \sin ^{2}\frac{%
\theta _{A}}{2}\cos ^{2}\frac{\theta _{B}}{2}  \notag \\
&&-\frac{1}{4}\left[ 2+e^{-2\gamma t}\right] \sin \theta _{A}\sin \theta
_{B}\sin \left( \phi _{A}+\phi _{B}\right)  \notag \\
&&-\frac{5}{4}\sin \theta _{A}\sin \theta _{B}\sin \left( \phi _{A}-\phi
_{B}\right)  \label{payoff b}
\end{eqnarray}%
For this case $Q\otimes Q$ with $Q\longrightarrow \left( \theta _{A},\phi
_{A},\theta _{B},\phi _{B}\right) =(0,\frac{\pi }{2},0,\frac{\pi }{2})$ is
the NE of the game with payoffs $\$_{A}(0,\frac{\pi }{2},0,\frac{\pi }{2}%
)=\$_{B}(0,\frac{\pi }{2},0,\frac{\pi }{2})=2+e^{-2\gamma t}$ \cite{eisert}.
For the analysis of this NE we apply the following NE conditions 
\begin{eqnarray}
\$_{A}(0,\frac{\pi }{2},0,\frac{\pi }{2})-\$_{A}(\theta _{A},\phi _{A},0,%
\frac{\pi }{2}) &\geq &0  \notag \\
\$_{B}(0,\frac{\pi }{2},0,\frac{\pi }{2})-\$_{B}(0,\frac{\pi }{2},\theta
_{B},\phi _{B}) &\geq &0.  \label{NE}
\end{eqnarray}%
With the help of Eqs. (\ref{payoff a}, \ref{payoff b}) the above
inequalities give%
\begin{equation}
7e^{-2\gamma t}+\left[ \left( 2\cos 2\phi _{i}-5\right) e^{-2\gamma t}+1%
\right] \cos ^{2}\frac{\theta _{i}}{2}-1\geq 0  \label{ineq-pi/2}
\end{equation}%
where $i=A,B.$ These inequalities are satisfied for all $\phi ^{\prime }s$
and $\theta ^{\prime }s$ if $e^{-2\gamma t}\geq \frac{1}{7}$. With
increasing $\gamma t$ when $e^{-2\gamma t}<\frac{1}{7}$ then the
inequalities (\ref{ineq-pi/2}) are not satisfied and $Q\otimes Q$ does not
remain NE but $Q\otimes D$ and $D\otimes Q$\ appear as NE in the game. For $%
Q\otimes D$ as NE we require 
\begin{eqnarray}
\$_{A}(0,\frac{\pi }{2},\pi ,0)-\$_{A}(\theta _{A},\phi _{A},\pi ,0) &\geq &0
\notag \\
\$_{B}(0,\frac{\pi }{2},\pi ,0)-\$_{B}(0,\frac{\pi }{2},\theta _{B},\phi
_{B}) &\geq &0.  \label{qd}
\end{eqnarray}%
Using Eqs. (\ref{payoff a}, \ref{payoff b}) inequalities (\ref{qd}) become%
\begin{eqnarray}
\left[ 7-\left( 2-5\cos 2\phi _{A}\right) \cos ^{2}\frac{\theta _{A}}{2}%
\right] e^{-2\gamma t}+\sin ^{2}\frac{\theta _{A}}{2} &\geq &0 \\
\left[ 1-\left( 5+2\cos 2\phi _{B}\right) e^{-2\gamma t}\right] \cos ^{2}%
\frac{\theta _{B}}{2} &\geq &0
\end{eqnarray}%
The above inequalities are satisfied for all $\theta ^{\prime }s$ and $\phi
^{\prime }s$ subject to the condition $0\leq e^{-2\gamma t}\leq \frac{1}{7}.$
It shows that $Q\otimes D$ remains NE for all values of $\gamma t$ for which 
$0\leq e^{-2\gamma t}\leq \frac{1}{7}.$ By similar reasoning $D\otimes Q$
can be proved NE for all values of $\gamma t$ for which $0\leq e^{-2\gamma
t}\leq \frac{1}{7}.$ However when $e^{-2\gamma t}\longrightarrow 0$ then
besides $Q\otimes D$ and $D\otimes Q$ the strategy pairs $C\otimes D$ and $%
D\otimes C$ also become NE. At this stage we have four NE in the game. The
NE conditions for $\left( C,D\right) $ are 
\begin{eqnarray}
\$_{A}(0,0,\pi ,0)-\$_{A}(\theta _{A},\phi _{A},\pi ,0) &\geq &0  \notag \\
\$_{B}(0,0,\pi ,0)-\$_{B}(0,0,\theta _{B},\phi _{B}) &\geq &0.
\end{eqnarray}%
With the help of Eqs. (\ref{payoff a}, \ref{payoff b}) the above
inequalities become%
\begin{eqnarray}
\sin ^{2}\frac{\theta _{A}}{2}-\left[ 3+\left( 2-5\cos 2\phi _{A}\right)
\cos ^{2}\frac{\theta _{A}}{2}\right] e^{-2\gamma t} &\geq &0  \notag \\
\left[ 1+\left( 5-2\cos 2\phi _{B}\right) e^{-2\gamma t}\right] \cos ^{2}%
\frac{\theta _{B}}{2} &\geq &0.  \label{cd}
\end{eqnarray}%
The inequalities (\ref{cd}) are satisfied for all $\theta ^{\prime }s$ and $%
\phi ^{\prime }s$ only if $e^{-2\gamma t}$ $\longrightarrow 0$. Similarly it
can be proved that $D\otimes C$ is NE when $e^{-2\gamma t}\longrightarrow 0.$

\subsection{Product Measurement}

For the measurement in product basis i.e. $\delta =0$, the payoffs given in
Eq. (\ref{payoff-pd-a}) and (\ref{payoff-pd-b}) for player $i=A$\ or $B$
become 
\begin{eqnarray}
\$_{i}(\theta _{A},\phi _{A},\theta _{B},\phi _{B}) &=&2-\cos ^{2}\frac{1}{2}%
\theta _{A}\cos ^{2}\frac{1}{2}\theta _{B}+\frac{1}{2}\cos ^{2}\frac{1}{2}%
\theta _{B}+  \notag \\
&&\frac{1}{2}\cos ^{2}\frac{1}{2}\theta _{A}-\frac{1}{4}e^{-2\gamma t}\sin
\theta _{A}\sin \theta _{B}\sin \left( \phi _{A}+\phi _{B}\right) .
\label{product}
\end{eqnarray}%
In this case the strategy pair $D\otimes C$ and $C\otimes D$ are two NE of
game with payoffs $\left( \frac{5}{2},\frac{5}{2}\right) $. For $D\otimes C$
the NE conditions $\$_{A}(\pi ,0)-\$_{A}(\theta _{A},0)\geq 0$ and $%
\$_{B}(\pi ,0)-\$_{B}(\pi ,\theta _{B})\geq 0$ give $\cos ^{2}\frac{\theta
_{A}}{2}\geq 0$ and $\sin ^{2}\frac{\theta _{B}}{2}\geq 0$ respectively.\
Both these conditions are always satisfied and are independent of
decoherence effects. Therefore $D\otimes C$ is NE for all values of $\gamma
t.$ Similarly it can be proved that $C\otimes D$ is a NE with same
properties. It is to be noted that classical game can not be reproduced in
this case. It is due to the fact that when a quantum game starts with an
entangled state of the form $\psi _{in}=\cos \frac{\xi }{2}\left\vert
00\right\rangle +i\sin \frac{\xi }{2}\left\vert 11\right\rangle $ and
measurement is performed in product basis then Marinatto and Weber
quantization scheme results \cite{marinatto}. In this scheme the classical
results can be reproduced with an unentangled initial quantum state with $%
\xi =0.$ But in our case game starts with a maximally entangled state that
has decohered by collective dephasing with dephasing rate $\gamma $ (see \ref%
{after dephasing}). At $t=0$\ (i.e. $e^{-2\gamma t}=1$)\ the initial state
of game is maximally entangled state and for all $t>0$ the initial state
becomes mixed. No value of $\gamma t$\ can be found that can transform Eq. ( %
\ref{after dephasing}) to a state that is required to reproduce the
classical game. This highlights the fact that initial quantum state plays a
crucial role in the solution of quantum games \cite%
{flitney,azhar,azhar1,azhar2,jiang,rosero}.

\subsection{Three Parameters Set of Strategies}

For three parameters set of strategies the players are equipped with the
unitary operators of the form 
\begin{equation}
\hat{U}(\theta ,\phi ,\psi )=\left[ 
\begin{array}{cc}
e^{i\phi }\cos \frac{\theta }{2} & ie^{i\psi }\sin \frac{\theta }{2} \\ 
ie^{-i\psi }\sin \frac{\theta }{2} & e^{-i\phi }\cos \frac{\theta }{2}%
\end{array}%
\right] ,
\end{equation}%
and the payoffs of players for initial state given by Eq. (\ref{after
dephasing})\ become

\begin{eqnarray}
\$_{A}(\theta _{A},\phi _{A},\psi _{A},\theta _{B},\phi _{B},\psi _{B}) &=& 
\left[ 2+e^{-2\gamma t}\cos 2\left( \phi _{A}+\phi _{B}\right) \sin \delta %
\right] \cos ^{2}\frac{\theta _{A}}{2}\cos ^{2}\frac{\theta _{B}}{2}  \notag
\\
&&+\left[ 2-e^{-2\gamma t}\cos 2\left( \psi _{A}+\psi _{B}\right) \sin
\delta \right] \sin ^{2}\frac{\theta _{A}}{2}\sin ^{2}\frac{\theta _{B}}{2} 
\notag \\
&&+\frac{5}{2}\left[ 1-e^{-2\gamma t}\cos 2\left( \phi _{A}-\psi _{B}\right)
\sin \delta \right] \cos ^{2}\frac{\theta _{A}}{2}\sin ^{2}\frac{\theta _{B}%
}{2}  \notag \\
&&+\frac{5}{2}\left[ 1+e^{-2\gamma t}\cos 2\left( \phi _{B}-\psi _{A}\right)
\sin \delta \right] \sin ^{2}\frac{\theta _{A}}{2}\cos ^{2}\frac{\theta _{B}%
}{2}  \notag \\
&&+\frac{1}{4}\sin \theta _{A}\sin \theta _{B}\left[ e^{-2\gamma t}\sin
\left( \phi _{A}+\phi _{B}-\psi _{A}-\psi _{B}\right) \right.  \notag \\
&&+\sin \delta \left\{ 2\sin \left( \phi _{A}+\phi _{B}+\psi _{A}+\psi
_{B}\right) \right.  \notag \\
&&-\left. \left. 5\sin \left( \phi _{A}-\phi _{B}+\psi _{A}-\psi _{B}\right)
\right\} \right]  \label{three-parameter-a}
\end{eqnarray}%
$\allowbreak $

\begin{eqnarray}
\$_{B}(\theta _{A},\phi _{A},\psi _{A},\theta _{B},\phi _{B},\psi _{B}) &=& 
\left[ 2+e^{-2\gamma t}\cos 2\left( \phi _{A}+\phi _{B}\right) \sin \delta %
\right] \cos ^{2}\frac{\theta _{A}}{2}\cos ^{2}\frac{\theta _{B}}{2}  \notag
\\
&&+\left[ 2-e^{-2\gamma t}\cos 2\left( \psi _{A}+\psi _{B}\right) \sin
\delta \right] \sin ^{2}\frac{\theta _{A}}{2}\sin ^{2}\frac{\theta _{B}}{2} 
\notag \\
&&+\frac{5}{2}\left[ 1+e^{-2\gamma t}\cos 2\left( \phi _{A}-\psi _{B}\right)
\sin \delta \right] \cos ^{2}\frac{\theta _{A}}{2}\sin ^{2}\frac{\theta _{B}%
}{2}  \notag \\
&&+\frac{5}{2}\left[ 1-e^{-2\gamma t}\cos 2\left( \phi _{B}-\psi _{A}\right)
\sin \delta \right] \sin ^{2}\frac{\theta _{A}}{2}\cos ^{2}\frac{\theta _{B}%
}{2}  \notag \\
&&+\frac{1}{4}\sin \theta _{A}\sin \theta _{B}\left[ e^{-2\gamma t}\sin
\left( \phi _{A}+\phi _{B}-\psi _{A}-\psi _{B}\right) \right.  \notag \\
&&+\sin \delta \left\{ 2\sin \left( \phi _{A}+\phi _{B}+\psi _{A}+\psi
_{B}\right) \right.  \notag \\
&&+\left. \left. 5\sin \left( \phi _{A}-\phi _{B}+\psi _{A}-\psi _{B}\right)
\right\} \right]  \label{three-parameter-b}
\end{eqnarray}%
$\allowbreak $

In this case there is no pure strategy NE because for every strategy of
Alice, Bob also has a counter strategy. However there can be a mixed
strategy (non unique) NE \cite{eisert1,benjamin}. The NE\ occurs when Alice
chooses the operators $A_{1}=\left( \theta _{A}=0,\phi _{A}=0,\psi _{A}=\psi
\right) $ and $A_{2}=\left( \theta _{A}=0,\phi _{A}=\frac{\pi }{2},\psi
_{A}=\psi \right) $ with equal probability and Bob chooses $B_{1}=\left(
\theta _{B}=\pi ,\phi _{B}=\phi ,\psi _{B}=0\right) $ and $B_{2}=\left(
\theta _{B}=\pi ,\phi _{B}=\phi ,\psi _{B}=\frac{\pi }{2}\right) $ with
equal probability. The payoffs corresponding to each pair of strategy are 
\begin{eqnarray*}
\$_{A}\left( A_{1},B_{1}\right)  &=&\frac{5}{2}\left( 1+e^{-2\gamma t}\sin
\delta \right) \text{, \ \ \ }\$_{A}\left( A_{1},B_{2}\right) =\frac{5}{2}%
\left( 1-e^{-2\gamma t}\sin \delta \right)  \\
\$_{A}\left( A_{2},B_{1}\right)  &=&\frac{5}{2}\left( 1-e^{-2\gamma t}\sin
\delta \right) \text{, \ \ \ }\$_{A}\left( A_{2},B_{2}\right) =\frac{5}{2}%
\left( 1+e^{-2\gamma t}\sin \delta \right)  \\
\$_{B}\left( A_{1},B_{1}\right)  &=&\frac{5}{2}\left( 1-e^{-2\gamma t}\sin
\delta \right) \text{, \ \ \ }\$_{B}\left( A_{1},B_{2}\right) =\frac{5}{2}%
\left( 1+e^{-2\gamma t}\sin \delta \right)  \\
\$_{B}\left( A_{2},B_{1}\right)  &=&\frac{5}{2}\left( 1+e^{-2\gamma t}\sin
\delta \right) \text{, \ \ \ }\$_{A}\left( A_{2},B_{2}\right) =\frac{5}{2}%
\left( 1-e^{-2\gamma t}\sin \delta \right) .
\end{eqnarray*}%
and payoff for both players at NE is $\frac{5}{2}$. Although the payoffs
against an individual pair of strategy depends upon decoherence the average
payoff at NE remains independent of decoherence both for entangled ($\delta =%
\frac{\pi }{2}$) and product measurements ($\delta =0$).

\section{\label{con}conclusion}

We quantized PD using generalized quantization scheme for the case when
initial state of game is maximally entangled state that has decohered by
collective dephasing of dephasing rate $\gamma $. In generalized
quantization scheme we have the options of performing measurement in
entangled as well as in product basis. In the case of entangled basis
measurements when the players are allowed to play two parameters set of
strategies then $Q\otimes Q$ is NE for $e^{-2\gamma t}>\frac{1}{7}$. With
increasing time when $e^{-2\gamma t}<\frac{1}{7}$ then $Q\otimes Q$
disappears as NE and two new NE $Q\otimes D$\ and $D\otimes Q$ appear
simultaneously. $Q\otimes D$ and $D\otimes Q$ remain NE for $0\leq
e^{-2\gamma t}\leq \frac{1}{7}$ however, when $e^{-2\gamma t}\rightarrow 0$
then $C\otimes D$\ and $D\otimes C$ also become NE resulting four NE in the
game.\emph{\ }On the other hand for the measurement in product basis there
are two NE $C\otimes D$\ and $D\otimes C$ in the game which are not affected
by decoherence. For three parameters set of strategies there is mixed
strategy NE that is also independent of decoherence for measurement in
entangled as well as in product basis.

Du \textit{et al} \cite{du} also generalized Eisert quantization scheme for
PD to study the effects of entanglement on NE by taking an initial quantum
state of the form 
\begin{equation}
\psi _{in}=\cos \frac{\xi }{2}\left\vert 00\right\rangle +i\sin \frac{\xi }{2%
}\left\vert 11\right\rangle   \label{du state}
\end{equation}%
where $\xi \in \left[ 0,\frac{\pi }{2}\right] $ is the measure of
entanglement. They showed that, for two parameters set of strategies, $%
Q\otimes Q$ is only a NE of PD when the entanglement of initial quantum
state is greater than certain threshold value $\xi _{th1}=.685.$ When
entanglement of initial state becomes less than this threshold value then
two new NE $Q\otimes D$ and $D\otimes Q$ appear in the game. This\ feature
of the game holds up to $\xi _{th2}=.464$. For any entanglement in the range 
$0\leq \xi \leq .464$ the game shows the classical behavior with $D\otimes D$%
\ as the NE. Our results are consistence with Du \textit{et al} results \cite%
{du} with two exceptions. First, the threshold value of entanglement
parameter for a particular NE is different. Second, at some minimum value of
entanglement Du \textit{et al} game behaves like a classical PD with $%
D\otimes D$ as the NE whereas in our case this feature is absent. We see
that it is due to the fact that in Du \textit{et al} \cite{du} scheme the
game starts from a maximally entangled pure state of the form (\ref{du state}%
)\ that remains pure for all values of $\xi \geq 0.$ But in our case the
game starts from an initial state of the form (\ref{after dephasing}) that
is maximally entangled pure state only at $\gamma t=0.$ For all $\gamma t>0$
it becomes mixed and for maximum decoherence (i.e. when $e^{-2\gamma
t}\rightarrow 0$) it transforms to $\frac{1}{2}\left( \left\vert
00\right\rangle \left\langle 00\right\vert +\left\vert 11\right\rangle
\left\langle 11\right\vert \right) $. As initial state plays an important
role in the solution of quantum games \cite{eisert,marinatto,azhariq} the
features of PD in our case are some what different than that of Du \textit{%
et al} \cite{du}.

\end{document}